\documentclass[aps,preprintnumbers,floatfix,article,amsmath,amssymb,floatfix,10pt,prd,superscriptaddress,nofootinbib]{revtex4-2}
\usepackage{bm}
\usepackage{amsfonts}
\usepackage{latexsym}
\usepackage{enumitem}
\usepackage{graphicx}
\usepackage{amsmath}
\usepackage{palatino}
\usepackage{mathpazo}
\usepackage{textcomp}
\usepackage{float}
\usepackage{booktabs}
\usepackage{dcolumn}
\usepackage{ragged2e}
\usepackage{hyperref}
\hypersetup{colorlinks,citecolor=magenta}
\hypersetup{colorlinks=true,linkcolor=red,filecolor=magenta,urlcolor=blue}
\usepackage{amsmath}
\usepackage{xcolor}
\usepackage{orcidlink}
\usepackage{epsfig}
\usepackage{caption}
\usepackage{subcaption}
\usepackage{commath}
\def\jnl@style{\it}
\def\aaref@jnl#1{{\jnl@style#1}}

\def\aaref@jnl#1{{\jnl@style#1}}

\def\aj{\aaref@jnl{AJ}}          
\def\apj{\aaref@jnl{ApJ}}         
\def\apjl{\aaref@jnl{ApJ}}        
\def\apjs{\aaref@jnl{ApJS}}        
\def\apss{\aaref@jnl{Ap\&SS}}       
\def\aap{\aaref@jnl{A\&A}}        
\def\aapr{\aaref@jnl{A\&A~Rev.}}     
\def\aaps{\aaref@jnl{A\&AS}}       
\def\mnras{\aaref@jnl{Mon.~Not.~Roy.~Astron.~Soc.}}       
\def\prd{\aaref@jnl{Phys.~Rev.~D}}    
\def\prc{\aaref@jnl{Phys.~Rev.~C}} 
\def\prl{\aaref@jnl{Phys.~Rev.~Lett.}}  
\def\qjras{\aaref@jnl{QJRAS}}       
\def\skytel{\aaref@jnl{S\&T}}       
\def\ssr{\aaref@jnl{Space~Sci.~Rev.}}   
\def\zap{\aaref@jnl{ZAp}}         
\def\nat{\aaref@jnl{Nature}}       
\def\aplett{\aaref@jnl{Astrophys.~Lett.}} 
\def\apspr{\aaref@jnl{Astrophys.~Space~Phys.~Res.}} 
\def\physrep{\aaref@jnl{Phys.~Rep.}}   
\def\physscr{\aaref@jnl{Phys.~Scr}}    
\def\commat{\aaref@jnl{Comm.~Math.~Phys.}}       
\def\science{\aaref@jnl{Science}}        
\def\cqg{\aaref@jnl{Classical Quant.~Grav.}}      
\def\jpcs{\aaref@jnl{JPCS}}                   
\def\ijmpd{\aaref@jnl{Int.~J.~Mod.~Phys.~D}}          
\def\grg{\aaref@jnl{Gen.~Relat.~Gravit.}}        
\def\rpp{\aaref@jnl{Rep.~Prog.~Phys.}}     
\def\npa{\aaref@jnl{Nucl.~Phys.~A}}    
\def\lrr{\aaref@jnl{Living Rev.~Rel.}}          
\def\jcap{\aaref@jnl{J.~Cosmology Astropart.~Phys.}}  
\def\rmp{\aaref@jnl{Rev.~Mod.~Phys.}}  
\def\epjc{\aaref@jnl{Eur.~Phys.~J.~C}} 
\def\plb{\aaref@jnl{~Phy.~Lett.~B}} 
\def\mpla{\aaref@jnl{Mod.~Phy.~Lett.~A}} 
\def\arxiv{\aaref@jnl{arxiv.org}}

\allowdisplaybreaks[1]
\renewcommand{\arraystretch}{1.1}
\begin{document}
\color{black}    
\title{\bf Baryon asymmetry constraints on Extended Symmetric Teleparallel Gravity}

\author{S. A. Narawade\orcidlink{0000-0002-8739-7412}}
\email{shubhamn2616@gmail.com}
\affiliation{Department of Mathematics,
Birla Institute of Technology and Science-Pilani, Hyderabad Campus,
Hyderabad-500078, India.}

\author{S.K. Tripathy \orcidlink{0000-0001-5154-2297}}
\email{tripathy\_sunil@rediffmail.com}
\affiliation{Department of Physics, Indira Gandhi Institute of Technology, Sarang, Dhenkanal, Odisha-759146, India.}

\author{Raghunath Patra}
\email{raghunathpatra09@gmail.com}
\affiliation{Department of Mathematics, Berhampur University, Berhampur, Odisha}

\author{B. Mishra\orcidlink{0000-0001-5527-3565}}
\email{bivu@hyderabad.bits-pilani.ac.in}
\affiliation{Department of Mathematics,
Birla Institute of Technology and Science-Pilani, Hyderabad Campus,
Hyderabad-500078, India.}

\begin{abstract}
In this paper, we have explored the observed matter-antimatter asymmetry in the Universe to constrain the model parameters in extended symmetric teleparallel gravity (STG) or $f(Q,T)$ gravity, where $Q$ be the nonmetricity and $T$ be the trace of energy momentum tensor. We have considered two functional forms of $f(Q,T)$ to find the baryon asymmetry to entropy ratio calculated at a decoupling temperature. Two different data sets namely Hubble data set and the Hubble+BAO+Pantheon data sets are used to constrain the scale factor and the constrained model is used to obtain the baryon asymmetry to entropy ratio. It is observed that, model constrained from the Hubble data set favour a narrow range of the $f(Q,T)$ gravity parameters to reproduce the observed baryon asymmetry.
\end{abstract}

\maketitle

\section{Introduction}
Relativistic quantum theory predicts a balance between the amount of matter and antimatter. However, measurements of Cosmic Microwave Background (CMB) \cite{Bennett03,Spergel03}, Big-Bang Nucleosynthesis (BBN) \cite{Burles01} and negligible radiation contribution from matter-antimatter annihilation \cite{Cohen98} show an excess of matter over antimatter in the Universe. In fact, locally the Universe is mostly dominated by matter. The process of generation of excess matter or baryons over anitmatter or antibaryons is baryogenesis. Essentially, Baryogenesis is an explanation of why matter is abundant in the universe while antimatter is scarce. This imbalance is the reason galaxies, stars, planets, and life exist. Baryogenesis is cosmologically significant because of its intimate role in the creation and evolution of the universe. It is imperative for understanding the large-scale structure and dynamics of the cosmos to understand the processes that lead to the prevalence of matter over antimatter. Secondly, baryogenesis provides insight into the origin of matter and contributes to our understanding of the fundamental forces and particles governing the universe because it violates fundamental symmetries in the early universe. A third reason for studying baryogenesis is that it is closely related to particle physics and high-energy physics, thereby providing insight into particle behavior at extreme levels, contributing to the refinement and testing of theories. Lastly, recent advances in experimental techniques, such as those employed in high-energy particle accelerators and astrophysical observations, offer the potential to test and constrain theories related to baryogenesis, providing an opportunity to highlight the current state of research in the field. Eventhough the observed baryon asymmetry is believed to be generated dynamically as the Universe expands and cools off, there is no specific reason proposed. According to the Sakharov principle the possible conditions for baryon asymmetry generation may be (i) the non-conserving interactions of baryon number (B), (ii) the violation of $\mathcal{C}$ and $\mathcal{C}P$ and (iii) departure from thermal equilibrium \cite{Sakharov67,Sakharov91}. The gravitational baryogenesis involves the presence of a CP violating interaction term and can also proceed in thermal equilibrium. The baryon asymmetry is usually defined as 
\begin{equation}
  \frac{n_b}{s}=\frac{n_B-n_{\Bar{B}}}{s}\simeq \frac{\eta}{7},
\end{equation}
where $n_B$ and $n_{\Bar{B}}$ are the baryon and antibaryon number densities respectively. $s=\frac{2\pi^2 g_{*s}}{45}\mathcal{T}^3$ is the entropy density in the radiation dominated era. $g_{*s}$ is the number of degrees of freedom for particles contributing to the entropy of the Universe and may be close to that in Standard model i.e $g_{*s}\simeq g_{*}=106$ \cite{Kolb90}. $\mathcal{T}$ is the temperature. The observed baryon asymmetry is $\frac{n_b}{s}=9.2_{-0.4}^{+0.6}\times 10^{-11}$ and $5.5\times 10^{-11} \leq \eta \leq 9.9\time 10^{-11}$ \cite{Lambiase2013}.

Modified theories of gravity are being proposed post-supernovae observation to address the late time cosmic acceleration issue. The gravitational action of these modified theories of gravity can be defined through the curvature, torsion and nonmetricity approaches. 
In an expanding Universe, the coupling between the curvature and the baryon number current dynamically breaks CPT, which leads to the baryon asymmetry. Gravitational baryogenesis and possible ways baryon asymmetry regeneration have been investigated in different modified gravity theories, for example, in $f(R)$ gravity \cite{Lambiase06,Lambiase07,Ramos17,Aghamohammadi18,Agrawal21a}, $f(R,T)$ gravity \cite{Nozari18}, $f(R, T, X)$ gravity \cite{Saleem22, Saleem23}, $f(R,P)$ gravity \cite{Rani23}, $f(T)$ gravity \cite{Oikonomou16,Bhattacharjee20}, $f(T, \Theta)$ gravity \cite{Rani22}, $f(R,G)$ gravity \cite{Odintsov16,Azhar21}, $f(G)$ gravity \cite{Chakraborty23} and so on.\\

In the nonmetricity approach, Nester and Yo \cite{Nester99} suggested the symmetric teleparallel gravity(STG), where the gravitational action is represented by the nonmetricity $Q$. Recently, the nonmetricity gravity developed into the coincident General Relativity of $f(Q)$ gravity \cite{Jimenez18}. Further Xu et al. \cite{Xu19} extended to $f(Q,T)$ gravity, where $T$ be the trace of energy momentum tensor. Of late, this nonmetricity based modified gravity theory has provided some promising results in different aspects of cosmology \cite{Pati21,Pati22,Pati23,Narawade23}. 

In $f(Q,T)$ gravity, the $\mathcal{C}P$ violating interaction reads as, 
\begin{equation}\label{eq:1}
\frac{1}{M^{2}_{*}}\int{d^{4}x}\sqrt{-g}(\partial_{\mu}(Q+T))J^{\mu},
\end{equation}
where, $M_{*}$ represents the cut-off scale of the order of reduced Planck mass, which characterizes the effective gravitational theory and $g$ be the metric determinant. In thermal equilibrium, $n_{b}$ is expressed as \cite{Davoudias04},
\begin{equation}\label{eq:2}
n_{b} = n_{B}- n_{\overline{B}} = \frac{g_{b}T^{3}}{6\pi^{2}}\left( \frac{\pi^{2}\mu_{B}}{\mathcal{T}}+ \left( \frac{\mu_{B}}{\mathcal{T}} \right)^{3}\right),
\end{equation}
where $g_{b}\simeq 1$ and $\mu_{B}$ respectively denote the intrinsic degree of freedom of baryons and the chemical potential. Also, $\mu_{B} = -\mu_{\overline{B}} \thicksim \pm\frac{\dot{Q} + \dot{T}}{M^{2}_{*}}$. Now, the corresponding baryon asymmetry to entropy ratio becomes, 
\begin{equation}\label{eq:3}
\frac{n_{b}}{s}\simeq \left.-\frac{15g_{b}}{4\pi^{2}g_{s}}\frac{\dot{Q}+\dot{T}}{M^{2}_{*}\mathcal{T}}\right\vert_{\mathcal{T_{D}}},
\end{equation}

and consequently

\begin{equation}\label{eq:3}
\eta \simeq \left.-\frac{105g_{b}}{4\pi^{2}g_{s}}\frac{\dot{Q}+\dot{T}}{M^{2}_{*}\mathcal{T}}\right\vert_{\mathcal{T_{D}}}.
\end{equation}
where $\mathcal{T_{D}}$ is the decoupling temperature. 

Assuming the thermal equilibrium to exist, the Universe evolves gradually whose energy density is proportional to its temperature as \cite{Burles01,Bennett03},
\begin{equation}\label{eq:13}
\rho = \frac{\pi}{30}g_{s}\mathcal{T}^{4}.
\end{equation}

In principle, for a given gravity theory, the decoupling temperature is obtained from the above expression after evaluating the energy density along with the slope of the non-metricity and that of the trace of energy-momentum tensor. Then these values are substituted to calculate the baryon asymmetry to entropy ratio. In this paper, we explore the observational baryon asymmetry to constrain the parameter of the modified $f(Q,T)$ gravity theory. We consider two functional forms for $f(Q,T)$, one linear in $T$ and the other quadratic in $T$ both having only one unknown parameter. The paper is organised as: 
In Sec-\ref{sec:II}, we present the field equations of STG or $f(Q,T)$ gravity. In Sec-\ref{sec:III}, the baryon asymmetry within gravitational baryogenesis are obtained in $f(Q,T)$ gravity for two different models and the model parameters are constrained from observed matter-antimatter imbalance in the Universe. In Sec-\ref{sec:IV} we present our conclusion.

\section{Extended STG Field Equations}\label{sec:II}
The action of $f(Q,T)$ gravity is defined with the matter Lagrangian  $\mathcal{L}_{ m}$ as \cite{Xu19},
\begin{equation}\label{eq:4}
S = \int \frac{1}{16\pi}f(Q, T)\sqrt{-g}~d^{4}x + \int \mathcal{L}_{m}\sqrt{-g}~d^{4}x.
\end{equation}
Here the nonmetricity tensor is,
\begin{equation}\label{eq:5}
 Q_{\beta \mu \nu} = \nabla_{\beta}g_{\mu \nu} = \frac{\partial g_{\mu \nu}}{\partial x^{\beta}}
\end{equation}
with its traces 
\begin{equation}\label{eq:6}
Q_{\beta} = Q^{~~\mu}_{\beta~~\mu}~~~~and~~~~\tilde{Q}^{\beta} = Q^{~~\beta \mu}_{\mu}.
\end{equation}

The field equations of $f(Q,T)$ gravity can be obtained by varying the action \eqref{eq:4} with respect to the metric tensor as,
\begin{equation}\label{eq:9}
\frac{2}{\sqrt{-g}}\nabla_{\alpha}(\sqrt{-g}f_{Q}P^{\alpha}_{~\mu \nu}) + \frac{1}{2}g_{\mu \nu}f-f_{T}(T_{\mu\nu}+\Theta_{\mu\nu}) + f_{Q}(P_{\mu \nu \beta}Q^{~~\alpha \beta}_{\nu} - 2Q_{\alpha \beta \mu}P^{\alpha \beta}_{~~~\nu}) = -8\pi T_{\mu \nu}.
\end{equation}
where $T_{\mu \nu} = -\frac{2}{\sqrt{-g}}\frac{\delta \sqrt{-g}\mathcal{L}_{m}}{\delta g^{\mu \nu}}$ and $\Theta_{\mu\nu} = \delta_{\mu\nu}p-2T_{\mu\nu}$. $f_Q$ and $f_{T}$ respectively represent the partial derivatives of $f(Q,T)$ with respect to $Q$ and $T$. $P^{\beta}_{~\mu \nu}$ is the superpotential defined as
\begin{equation}\label{eq:7}
P^{\beta}_{~\mu \nu} \equiv -\frac{1}{4}Q^{\beta}_{~\mu \nu} + \frac{1}{4}\left(Q^{~\beta}_{\mu~\nu} + Q^{~\beta}_{\nu~~\mu}\right) + \frac{1}{4}Q^{\beta}g_{\mu \nu} - \frac{1}{8}\left(2 \tilde{Q}^{\beta}g_{\mu \nu} + {\delta^{\beta}_{\mu}Q_{\nu} + \delta^{\beta}_{\nu}Q_{\mu}} \right).
\end{equation}

 For a flat Friedmann-Lima$\hat{i}$tre-Robertson-Walker metric, 
\begin{equation}\label{eq:10}
ds^2=-dt^2+a^2(t)(dx^2+dy^2+dz^2),
\end{equation}
the nonmetricity scalar becomes $Q = 6H^{2}$, where $H=\frac{\dot{a}}{a}$ is the Hubble parameter and $a(t)$ is the scale factor. We may assume a perfect fluid distribution with $T_{\mu}^{\nu}=diag(-\rho, p, p, p)$ and $\Theta_{\mu}^{\nu}=diag(2\rho+p, -p, -p, -p)$ for which the field equations of $f(Q,T)$ gravity become \cite{Xu19,Pati21},
\begin{eqnarray}
\frac{f}{2} - 6f_QH^{2} -\frac{2\kappa}{1+\kappa}\dot{\xi} &=& 8\pi\rho,\label{eq:11}\\
\frac{f}{2}-6f_QH^{2} - 2\dot{\xi} &=&-8\pi p,\label{eq:12}
\end{eqnarray}
where, $8\pi\kappa = f_{T}$,$~~$ $\xi = f_{Q}H$ and we have an evolution equation for the Hubble function as $\dot{\xi} = 4\pi (\rho + p)(1+\kappa)$. 

Viable and stable cosmological models can be constructed using different functional forms of $f(Q,T)$. The parameters of the $f(Q,T)$ gravity models are usually constrained from different sets of cosmological data and then the constructed models are tested through a dynamical stability analysis to obtain stable regions. In the present work, we wish to consider two different models of $f(Q,T)$ gravity through a suitable choice of the functional $f(Q,T)$ that depends on certain arbitrary parameters which we may constrain from the gravitational baryogenesis and the observational baryon asymmetry to entropy ratio.

\section{Constraining Extended STG from Gravitational Baryogenesis}\label{sec:III}

We discuss here two models through the choice of the functional forms namely $f(Q,T)= -Q^{m}+ T$ and $f(Q,T) = -Q - \lambda T^{2}$. The choice is primarily motivated to use a linear form and a quadratic form of the trace of the energy-momentum tensor. Also, we restrict the models to have only one free positive parameter $m$ or $\lambda$, as the case may be, to be constrained from the baryon asymmetry. A hybrid scale factor (HSF) may be used to obtain the Hubble parameter $H = H_{0}\left(\alpha + \frac{\beta}{t}\right)$ with $H_0$, $\alpha$ and $\beta$ as constants. The constants have been constrained in a recent work \cite{Narawade23} using the Hubble and Hubble+BAO+Pantheon data sets and are given in TABLE-I.
\begin{table*}[b]
\begin{center}
\renewcommand{\arraystretch}{1.5}
\begin{tabular}{l c c c c c c c c c}
\hline 
Data & $H_{0}$ & $\alpha$ & $\beta$ \\
\hline
$Priors$ & $(60,80)$ & $(0,10)$ & $(0,10)$ \\
$Hubble$  & $67.7^{+1.7}_{-1.7}$ & $0.57^{+0.14}_{-0.13}$ & $0.38^{+0.12}_{-0.13}$ \\
$Hubble+BAO+Pantheon$  & $67.2^{+1.2}_{-1.2}$ & $0.603^{+0.028}_{-0.030}$ & $0.355^{+0.044}_{-0.043}$ \\
\hline
\end{tabular}
\caption{Best-fit values of parameter space using $Hubble$ and $Hubble+BAO+Pantheon$ data.}
\label{tab}
\end{center}
\end{table*}

\begin{enumerate}
\item {\textit{MODEL I: $f(Q,T)= -Q^{m}+ T$}}\\
The $f(Q,T)$ cosmological model with a linear term of the trace of the energy-momentum tensor, we have $f_{Q} = -mQ^{m-1}$, $~~$$8\pi\kappa=1$ and $\dot{\xi}=f_{Q}(2m-1)\dot{H}$. 

The energy density \eqref{eq:11} and pressure \eqref{eq:12} for this model become,
\begin{eqnarray}
\rho &=& \frac{(2m-1)Q^{m}+2\dot{\xi}\kappa}{4(4\pi+1)}\label{eq:14},\\
p &=& \frac{2\dot{\xi}(2+3\kappa)-(2m-1)Q^{m}}{4(4\pi+1)}.\label{eq:15}
\end{eqnarray}

The first order derivative of the nonmetricty $Q$ and trace $T=3p-\rho$ in cosmic time $t$ are
\begin{eqnarray}
\dot{Q} &=& -\frac{12\beta H^{2}_{0}(\beta+\alpha t)}{t^3},\label{eq:16}\\
\dot{T}&=& \frac{- 2^{m-2} 3^{m-1}\beta m(2 m-1)H_0^{2m-1} (\beta +\alpha t)^{2m-3} \left[\beta ((2\kappa +3) m-3 \beta H_{0})+\alpha t (-6 \beta H_{0}+3\kappa +2)-3 \alpha ^2 H_{0} t^2\right]}{\pi (2\kappa +1) t^{2m+1} }.\label{eq:17}
\end{eqnarray}
Substituting eqns. \eqref{eq:13}, \eqref{eq:14}, \eqref{eq:16} and \eqref{eq:17} in eqn. \eqref{eq:3}, we obtain the baryon asymmetry to entropy ratio as
\begin{equation}\label{eq:18.1}
\frac{n_{b}}{s}\simeq -\frac{15g_{b}}{4\pi^{2}g_{s}M^{2}_{*}}
\left(
\frac{\pi ^{3/4} \beta \left(\frac{-6^{m} m (2 m-1) t^2 \left(H^{2}_{0} \left(\alpha +\frac{\beta }{t}\right)^2\right)^m \left(\beta ((2\kappa +3) m-3 \beta H_{0})+\alpha t (-6 \beta H_{0}+3\kappa +2)-3 \alpha ^{2}H_{0} t^2\right)}{2 \pi\kappa +\pi }-144H^{3}_{0} (\beta +\alpha t)^4\right)}{6 \sqrt[4]{10}H_{0} t^3 (\beta +\alpha t)^3 \sqrt[4]{\frac{H_{0} 6^m (2 m-1) \left(\frac{H^{2}_{0} (\beta +\alpha t)^2}{t^2}\right)^{m-1} \left(3H_{0} (\beta +\alpha t)^2+\beta\kappa m\right)}{t^2 (2g_{s}\kappa +g_{s})}}}
\right).
\end{equation}

and 

\begin{equation}\label{eq:18}
\eta\simeq -\frac{105g_{b}}{4\pi^{2}g_{s}M^{2}_{*}}
\left(
\frac{\pi ^{3/4} \beta \left(\frac{-6^{m} m (2 m-1) t^2 \left(H^{2}_{0} \left(\alpha +\frac{\beta }{t}\right)^2\right)^m \left(\beta ((2\kappa +3) m-3 \beta H_{0})+\alpha t (-6 \beta H_{0}+3\kappa +2)-3 \alpha ^{2}H_{0} t^2\right)}{2 \pi\kappa +\pi }-144H^{3}_{0} (\beta +\alpha t)^4\right)}{6 \sqrt[4]{10}H_{0} t^3 (\beta +\alpha t)^3 \sqrt[4]{\frac{H_{0} 6^m (2 m-1) \left(\frac{H^{2}_{0} (\beta +\alpha t)^2}{t^2}\right)^{m-1} \left(3H_{0} (\beta +\alpha t)^2+\beta\kappa m\right)}{t^2 (2g_{s}\kappa +g_{s})}}}
\right).
\end{equation}

\begin{figure}[H]
\includegraphics[scale=0.5]{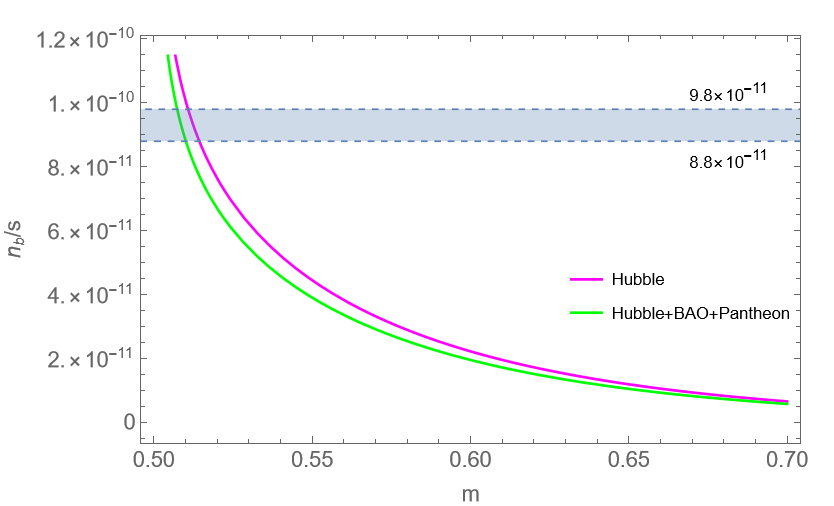}
\includegraphics[scale=0.5]{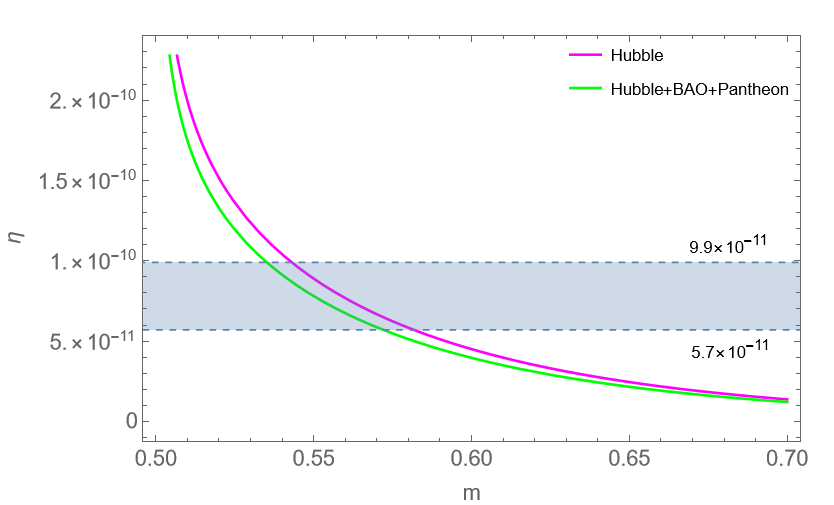}
 \caption{$\frac{n_{b}}{s}$ vs $m$ (left panel) and $\eta$ vs $m$ (right panel) with $g_{s}\simeq 106 $, $g_{b}\simeq 1$, $M_{*}= 10^{16}$ and $t = 10^{-7}$. The shaded bands represent the corresponding observational baryon asymmetry.}\label{FIG.1}
\end{figure}

The baryon asymmetry to entropy ratio $\frac{n_{b}}{s}$ (left panel) and $\eta$ (right panel) are shown as the function of the model parameter $m$ in [FIG.\ref{FIG.1}]. The shaded bands in the figures show the respective observational constraints on $\frac{n_{b}}{s}$ (left panel) and $\eta$(right panel). Corresponding to the two different data sets constraints on the scale factor parameters we have two different curves for these quantities. These curves show that the baryon asymmetry to entropy ratio and $\eta$ decrease with an increase in $m$ and pass through the observational band for some specific values of $m$. For the parameters obtained from Hubble data sets (pink curve), while the $\frac{n_b}{s}$ value constrains $m$ in the range $0.511\leq m \leq 0.515$, the $\eta$ value constrains in the range $0.543\leq m \leq 0.583$. For the parameters obtained from the combined dataset, $\frac{n_b}{s}$ value constrains $m$ in the range $0.508\leq m \leq 0.51$, the $\eta$ value constrains in the range $0.535\leq m\leq 0.572$. More or less, the baryon asymmetry favors a functional $f(Q,T)$ having a square root term of $Q$ with a linear trace term.

  \item {\textit{MODEL II: $f(Q,T) = -Q - \lambda T^{2}$}}

Here we wish to obtain the baryon asymmetry to entropy ratio and $\eta$ with a functional involving the squared trace. Further, we will treat the cosmological matter from the beginning as dust in order to keep the equation of state simple, i.e. $p = 0$. The trace becomes, $T = -\rho$, $f_{Q} = -1$ and $8\pi\kappa = -2\lambda T = 2\lambda \rho(t)$. 

From eqns. \eqref{eq:11} and \eqref{eq:12}, one may get
\begin{equation}\label{eq:23}
\rho = \frac{-\lambda\rho^{2}+6H^{2}}{16\pi(1+2\lambda\rho)}
\end{equation}
which may be solved to obtain,
\begin{equation}\label{eq:24}
\rho = \frac{-8\pi+\sqrt{64\pi ^{2}+6\lambda H^{2}(1+32\pi)}}{\lambda (1+32\pi)}
\end{equation}
and subsequently, we find
\begin{equation}\label{eq:25}
T = \left[ \frac{30}{g_{s}\pi} \left(\frac{-8\pi+\sqrt{64\pi ^{2}+6\lambda H^{2}(1+32\pi)}}{\lambda (1+32\pi)}\right)\right]^{\frac{1}{4}}.
\end{equation}
With an algebraic manipulation of eqns. \eqref{eq:3},\eqref{eq:13}, \eqref{eq:16}, \eqref{eq:24} and \eqref{eq:25}, the baryon asymmetry to entropy ratio is obtianed as
\begin{equation}\label{eq:26}
\frac{n_{b}}{s}\simeq -\frac{15g_{b}}{4\pi^{2}g_{s}M^{2}_{*}}
\left(
\frac{\sqrt{\pi } \sqrt[4]{\frac{1}{30} (1+32 \pi )} \left(\frac{3 \beta H^{2}_{0}(\beta +\alpha t)}{t^3 \sqrt{\frac{3 (1+32 \pi )H^{2}_{0}\lambda (\beta +\alpha t)^2}{2 t^2}+16 \pi ^2}}-\frac{12 \beta H^{2}_{0} (\beta +\alpha t)}{t^3}\right)}{\sqrt[4]{\frac{\sqrt{6 (1+32 \pi )H^{2}_{0} \lambda \left(\alpha +\frac{\beta }{t}\right)^2+64 \pi ^2}-8 \pi }{g_{s}\lambda}}}
\right),
\end{equation}

and 
\begin{equation}\label{eq:26}
\eta= -\frac{105g_{b}}{4\pi^{2}g_{s}M^{2}_{*}}
\left(
\frac{\sqrt{\pi } \sqrt[4]{\frac{1}{30} (1+32 \pi )} \left(\frac{3 \beta H^{2}_{0}(\beta +\alpha t)}{t^3 \sqrt{\frac{3 (1+32 \pi )H^{2}_{0}\lambda (\beta +\alpha t)^2}{2 t^2}+16 \pi ^2}}-\frac{12 \beta H^{2}_{0} (\beta +\alpha t)}{t^3}\right)}{\sqrt[4]{\frac{\sqrt{6 (1+32 \pi )H^{2}_{0} \lambda \left(\alpha +\frac{\beta }{t}\right)^2+64 \pi ^2}-8 \pi }{g_{s}\lambda}}}
\right).
\end{equation}

Fig. \ref{FIG.2} [left panel] represents the ratio of baryon asymmetry to entropy with respect to model parameter $\lambda$. In the right panel, $\eta$ is shown. The observational values of these quantities are shown as shaded bands in the figures. The $\frac{n_b}{s}$ and $\eta$ increase with an increase in the value of $\lambda$. The respective curves for the parameter sets constrained using the Hubble+BAO+Pantheon data sets remain below that of the Hubble data sets. The constraints obtained from $\frac{n_b}{s}$ are $0.425 \leq \lambda \leq 1.0$ (Hubble) and $1.23 \leq \lambda \leq 2.9$ (Hubble+BAO+Pantheon). The baryon asymmetry $\eta$ constrains $\lambda$ in the range $0.3 \leq \lambda \leq 1.71$ (Hubble) and $1.81 \leq \lambda \leq 4.29$ (Hubble+BAO+Pantheon). As in the previous case, for model parameters with the Hubble data set, the baryon asymmetry constrains the $f(Q,T)$ parameter in a narrow range.

\begin{figure}[H]
\includegraphics[scale=0.45]{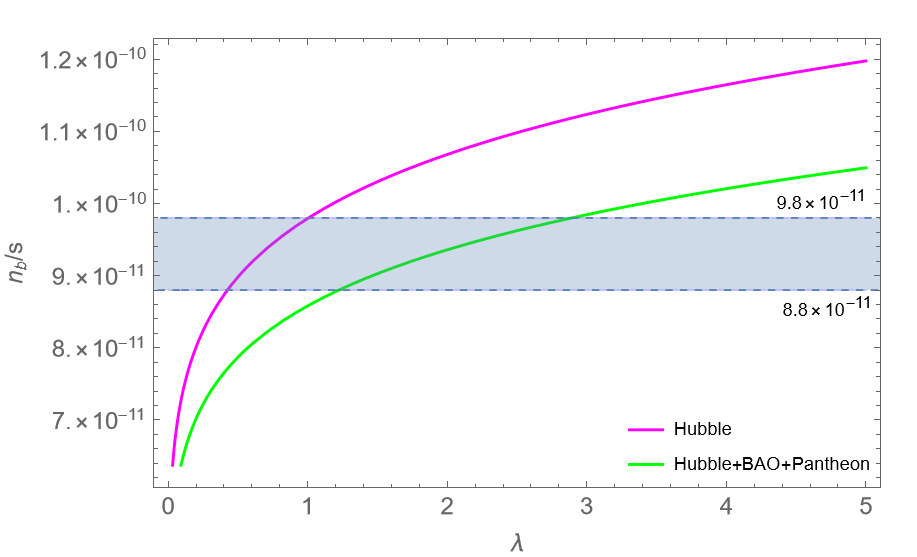}
\includegraphics[scale=0.45]{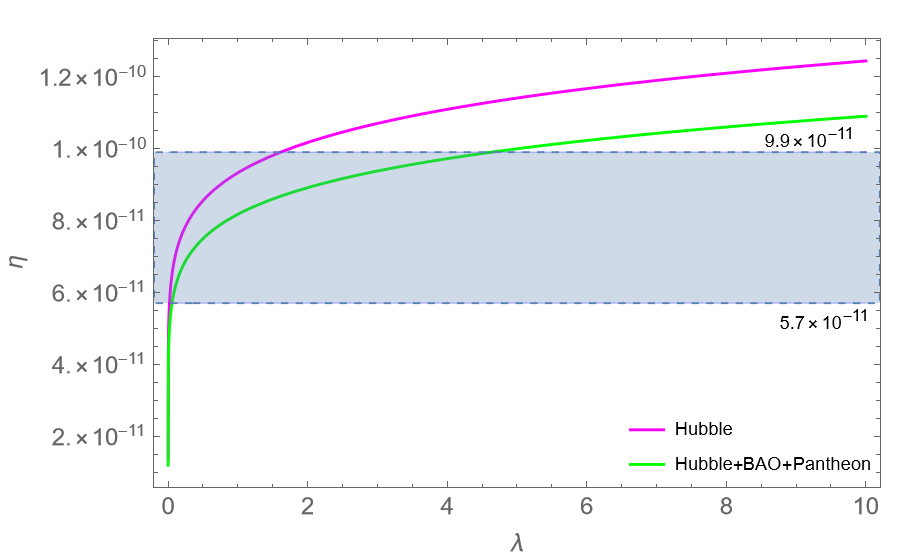}
\caption{$\frac{n_{b}}{s}$ vs $\lambda$ (left panel) and $\eta$ vs $\lambda$ (right panel) with $g_{s}\simeq 106 $, $g_{b}\simeq 1$, $M_{*}= 10^{16}$ and $t = 10^{-7}$. The shaded bands represent the corresponding observational baryon asymmetry.}\label{FIG.2}
\end{figure}

\end{enumerate}
 
\section{Conclusion}\label{sec:IV} 
We have calculated the baryon asymmetry of the Universe by considering the CP violation process in the framework of modified $f(Q, T)$ theory of gravity primarily developed to explain the late time cosmic speed up issue. In principle, it is possible to construct a good number of cosmological models by choosing different forms of the functional $f(Q,T)$ appearing in the gravitational action. We considered two different functional forms of the functional $f(Q,T)$, one linear in $T$ but have a power law for $Q$ and the other one is linear in $Q$ with a quadratic trace term. Both the models contain only one adjustable parameter. In the linear model, the baryon asymmetry to entropy ratio decreases with the parameter $m$ and in the quadratic case, it increases with an increase in $\lambda$. We explored the observed baryon asymmetry in nature to constrain the model parameters. Also, we observe that, the baryon asymmetry constrains the model from Hubble data set in a narrow range as compared to that from a combined data set. 

In summary, our study focuses in the insights into baryogenesis through the exploration of these two distinct models, emphasizing the role of model parameters in shaping the baryon asymmetry and providing constraints that align with observational data. The narrow parameter ranges favored by the baryon asymmetry further underscore the potential significance of specific functional forms in describing the trace of the energy-momentum tensor in cosmological models.

\section*{Acknowledgement} SKT and BM thanks IUCAA acknowledge the support of IUCAA, Pune (India) through the visiting associateship program. The authors are thankful to the anonymous referee for the suggestions to improve the quality of the paper.


\begin{thebibliography}{0}%
\makeatletter
\providecommand \@ifxundefined [1]{%
 \@ifx{#1\undefined}
}%
\providecommand \@ifnum [1]{%
 \ifnum #1\expandafter \@firstoftwo
 \else \expandafter \@secondoftwo
 \fi
}%
\providecommand \@ifx [1]{%
 \ifx #1\expandafter \@firstoftwo
 \else \expandafter \@secondoftwo
 \fi
}%
\providecommand \natexlab [1]{#1}%
\providecommand \enquote  [1]{``#1''}%
\providecommand \bibnamefont  [1]{#1}%
\providecommand \bibfnamefont [1]{#1}%
\providecommand \citenamefont [1]{#1}%
\providecommand \href@noop [0]{\@secondoftwo}%
\providecommand \href [0]{\begingroup \@sanitize@url \@href}%
\providecommand \@href[1]{\@@startlink{#1}\@@href}%
\providecommand \@@href[1]{\endgroup#1\@@endlink}%
\providecommand \@sanitize@url [0]{\catcode `\\12\catcode `\$12\catcode
  `\&12\catcode `\#12\catcode `\^12\catcode `\_12\catcode `\%12\relax}%
\providecommand \@@startlink[1]{}%
\providecommand \@@endlink[0]{}%
\providecommand \url  [0]{\begingroup\@sanitize@url \@url }%
\providecommand \@url [1]{\endgroup\@href {#1}{\urlprefix }}%
\providecommand \urlprefix  [0]{URL }%
\providecommand \Eprint [0]{\href }%
\providecommand \doibase [0]{https://doi.org/}%
\providecommand \selectlanguage [0]{\@gobble}%
\providecommand \bibinfo  [0]{\@secondoftwo}%
\providecommand \bibfield  [0]{\@secondoftwo}%
\providecommand \translation [1]{[#1]}%
\providecommand \BibitemOpen [0]{}%
\providecommand \bibitemStop [0]{}%
\providecommand \bibitemNoStop [0]{.\EOS\space}%
\providecommand \EOS [0]{\spacefactor3000\relax}%
\providecommand \BibitemShut  [1]{\csname bibitem#1\endcsname}%
\let\auto@bib@innerbib\@empty
\end{thebibliography}%


\begin{thebibliography}{99} 
\section*{References}

\bibitem{Spergel03} D.N. Spergel et al., \href{https://iopscience.iop.org/article/10.1086/377226}{\textit{Astrophys. J. Suppl. ser.}, \textbf{148}, 175 (2003)}.

\bibitem{Bennett03} C. L. Bennett et al., \href{https://iopscience.iop.org/article/10.1086/377253}{\textit{Astrophys. J. Suppl. ser.}, \textbf{148}, 1 (2003)}.

\bibitem{Burles01} S. Burles, K. M. Nollett, M. S. Turner, \href{https://doi.org/10.1103/PhysRevD.63.063512}{\textit{Phys. Rev. D}, \textbf{63}, 063512 (2001)}.

\bibitem{Cohen98} A. G. Cohen, A. De Rujula, S. L. Glashow, \href{https://iopscience.iop.org/article/10.1086/305328}{ \textit{Astrophys. J.}, \textbf{495}, 539 (1998).}

\bibitem{Sakharov67} A. D. Sakharov, \href{https://doi.org/10.1142/9789812815941_0013}{\textit{Zhurnal Eksperimental'noi i Teoreticheskoi Fiziki: Pis'ma v Redaktsiyu}, \textbf{5}, 32 (1967).}


\bibitem{Sakharov91} A. D. Sakharov, \href{https://doi.org/10.1070/PU1991v034n05ABEH002497}{\textit{Sov. Phys. Usp.}, \textbf{34}, 392 (1991).}

\bibitem{Kolb90} E.W. Kolb, M.S. Turner, \textit{The early Universe, Front. Phys.} \textbf{69}, 1 (1990).

\bibitem{Lambiase2013} G.Lambiase, S. Mohanty, A. R. Prasanna, \href{https://doi.org/10.1142/S0218271813300309}{\textit{Int. J. Mod. Phys. D} \textbf{22}, 1330030 (2013).}


\bibitem{Lambiase06} G. Lambiase, G. Scarpetta, \href{https://doi.org/10.1103/PhysRevD.74.087504}{\textit{Phys. Rev. D}, \textbf{74}, 087504 (2006).}

\bibitem{Lambiase07} G Lambiase, G Scarpetta, \href{https://iopscience.iop.org/article/10.1088/1742-6596/67/1/012055} {\textit{J. Phys.: Conf. Ser.}, \textbf{67}, 012055 (2007).}

\bibitem{Ramos17} M. P. L. P. Ramos et al., \href{https://doi.org/10.1103/PhysRevD.96.104024}{\textit{Phys. Rev. D}, \textbf{96}, 104024 (2017).}

\bibitem{Aghamohammadi18} A. Aghamohammadi, H. Hossienkhani, K. Saaidi, \href{https://doi.org/10.1142/S0217732318500724}{\textit{Mod. Phys. Lett. A}, \textbf{33}, 1850072 (2018).}

\bibitem{Agrawal21a} A. S. Agrawal et al., \href{https://doi.org/10.1016/j.cjph.2021.03.004}{\textit{C. J. Phys.}, \textbf{71}, 333 (2021).}

\bibitem{Nozari18} K. Nojari, F. Rajabi, \href{https://iopscience.iop.org/article/10.1088/0253-6102/70/4/451}{\textit{Commun. Theor. Phys.}, \textbf{70}, 451 (2018).}

\bibitem{Saleem22} R. Saleem, A. Saleem, \href{https://doi.org/10.1140/epjp/s13360-022-03181-w}{\textit{Eur. Phy. J. C.} \textbf{137}, 961 (2022).}

\bibitem{Saleem23} R. Saleem, A. Saleem, \href{https://dx.doi.org/10.1088/1402-4896/accbf2}{\textit{Phy. Scr.} \textbf{98}, 055021 (2023).}

\bibitem{Rani23} S. Rani, A. Javed, A. Jawad, \href{https://doi.org/10.1140/epjp/s13360-022-03557-y}{\textit{Eur. Phy. J. C.} \textbf{138}, 05 (2023).}

\bibitem{Oikonomou16} V. Oikonomou et al., \href{https://doi.org/10.1103/PhysRevD.94.124005}{\textit{Phys. Rev. D}, \textbf{94}, 124005 (2016).}

\bibitem{Bhattacharjee20} S. Bhattacharjee, \href{https://doi.org/10.1016/j.dark.2020.100612}{\textit{Phys. Dark Univ.}, \textbf{30}, 100612 (2020).}

\bibitem{Rani22} S. Rani, N. Azhar, A. Jawad, \href{https://doi.org/10.1142/S0217732322500560}{\textit{Mod. Phys. Lett. A}, \textbf{37}, 09 (2022).}

\bibitem{Odintsov16} S.D. Odintsov, V.K. Oikonomou, \href{https://doi.org/10.1016/j.physletb.2016.06.074}{\textit{Phys. Lett. B}, \textbf{760}, 259 (2016).}

\bibitem{Azhar21} N. Azhar et al., \href{https://doi.org/10.1016/j.dark.2021.100815}{\textit{Phys. Dark Universe}, \textbf{32}, 100815 (2021).}

\bibitem{Chakraborty23} G. Chakraborty, A. Beesham, S. Chattopadhyay, \href{https://doi.org/10.1142/S021988782350113X}{\textit{Int. J. Geom. Meth. Mod. Phys.}, \textbf{20}, 07 (2023).}

\bibitem{Nester99} J.M. Nester, H-J Yo, \href{}{\textit{Chin. J. Phys.}, \textbf{37}, 113 (1999).}

\bibitem{Jimenez18} J.B. Jimenez, L. Heisenberg, T. Koivisto, \href{https://doi.org/10.1103/PhysRevD.98.044048} {\textit{Phys. Rev. D}, \textbf{98}, 044048 (2018).}

\bibitem{Xu19} Y. Xu et al., \href{https://doi.org/10.1140/epjc/s10052-019-7207-4}{\textit{Eur. Phy. J. C.} \textbf{79}, 708 (2019).}

\bibitem{Pati21} L. Pati, B. Mishra, S.K. Tripathy, \href{https://doi.org/10.1088/1402-4896/ac0f92}{\textit{Phy. Scr.}, \textbf{96}, 105003 (2021).}

\bibitem{Pati22} L. Pati et al., \href{https://doi.org/10.1016/j.dark.2021.100925}{\textit{Phys. Dark Univ.}, \textbf{35}, 100925 (2022).}

\bibitem{Pati23} L. Pati et al., \textit{Eur. Phys. J. C}, In press, (2023). 

\bibitem{Narawade23} S. Narawade. M. Koussour, B.Mishra, \textit{Nucl. Phys. B}, In press (2023). 

\bibitem{Davoudias04} H. Davoudias et al., \href{https://doi.org/10.1103/PhysRevLett.93.201301}{\textit{Phys. Rev. Lett.}, \textbf{93}, 201301 (2004).}



\end{thebibliography}
\end{document}